\documentclass[journal=jacsat,manuscript=article]{achemso}

\usepackage[version=3]{mhchem} 
\usepackage{booktabs}

\newcommand{\onlinecite}[1]{\hspace{-1 ex} \nocite{#1}\citenum{#1}}
\author{August E. G. Mikkelsen}
\affiliation{Department of Energy Conversion and Storage, Technical University of Denmark, DK-2800 Kgs. Lyngby, Denmark}
\author{Henrik H. Kristoffersen}
\affiliation{Department of Chemistry, University of Copenhagen, Copenhagen 2100, Denmark}
\author{Jakob Schiøtz}
\affiliation{CAMD, Department of Physics, Technical University of Denmark, DK-2800 Kgs. Lyngby, Denmark}
\author{Tejs Vegge}
\affiliation{Department of Energy Conversion and Storage, Technical University of Denmark, DK-2800 Kgs. Lyngby, Denmark}
\altaffiliation{Corresponding author.}
\email{teve@dtu.dk}
\author{Heine A. Hansen}
\affiliation{Department of Energy Conversion and Storage, Technical University of Denmark, DK-2800 Kgs. Lyngby, Denmark}
\author{Karsten W. Jacobsen}
\affiliation{CAMD, Department of Physics, Technical University of Denmark, DK-2800 Kgs. Lyngby, Denmark}
\title{Room temperature structure and energetics of water-hydroxyl layers on Pt(111)}

\abbreviations{IR,NMR,UV}

\begin{document}


\begin{abstract}
The interactions between water and hydroxyl species on Pt(111) surfaces have been intensely investigated due to their importance to fuel cell electrocatalysis. Here we present a room temperature molecular dynamics study of their structure and energetics using an ensemble of neural network potentials, which allow us to obtain unprecedented statistical sampling. We first study the energetics of hydroxyl formation, where we find a near-linear adsorption energy profile, which exhibits a soft and gradual increase in the differential adsorption energy at high hydroxyl coverages. This is strikingly different from the predictions of the conventional bilayer model, which displays a kink at 1/3ML OH coverage indicating a sizeable jump in differential adsorption energy, but within the statistical uncertainty of previously reported ab initio molecular dynamics studies. We then analyze the structure of the interface, where we provide evidence for the water-OH/Pt(111) interface being hydrophobic at high hydroxyl coverages. We furthermore explain the observed adsorption energetics by analyzing the hydrogen bonding in the water-hydroxyl adlayers, where we argue that the increase in differential adsorption energy at high OH coverage can be explained by a reduction in the number of hydrogen bonds from the adsorbed water molecules to the hydroxyls.
\end{abstract}

\section{Introduction}
Water-metal interfaces are paramount to the field of electro-catalysis, as they constitute an indispensable component of modern electrochemical devices. Indeed, the energy conversion in e.g. PEM fuel cells relies on the oxygen reduction (ORR) and hydrogen oxidation (HOR) reactions, which take place at the interface between a metallic electrode and a water layer \cite{Debe}. Consequently much of recent developments in electro-catalysis have focused on understanding the catalytic properties of water-metal interfaces and how they depend on e.g. the choice of electrode material \cite{Norskov, Greeley, Stamenkovic}. \newline
\indent In the context of ORR and HOR, platinum is widely used as a catalyst as it is known to exhibit good electrocatalytic performance \cite{Debe, Norskov1, Durst}. Because of this the water-Pt interface is often considered the prototypical system for theoretical and computational models, and it has been studied in great detail at different levels of theory ranging from classical force fields \cite{Forcefield, Forcefield1} to the ab initio level \cite{Rossmeisl, Skulason, Tripkovic, Tripkovic1, CalleVallejo2014}. While significant insight has been gained there are still open questions, which remain to be adressed. In particular modelling the effects of the liquid water layer at ambient conditions has proved a major challenge, as finite temperature simulations based on ab initio molecular dynamics (AIMD) are limited to prohibitively small system sizes and short time scales. This can be problematic when modelling room temperature properties of the interface such as, e.g., the dynamic structure of the interfacial water layer \cite{Gross, Gross1, Gross2}. \newline  
\indent In a recent publication \cite{Mikkelsen2021}, we have discussed the above issue in detail and presented a possible solution based on fitting an ensemble of neural network potentials (NNPs). Using this approach, we demonstrated that it was possible to bridge the problems of obtaining high accuracy versus proper thermal sampling for the problem of determining the room temperature structure of the water-Pt(111) interface. \newline
\indent In this paper, we extend the methodology of our previous work \cite{Mikkelsen2021} to model the energetics of hydroxyl formation. Concretely, our main focus is the following OH adsorption reaction on a crystalline Pt(111) surface:
\begin{align}
 \text{H}_{2}\text{O}(\text{l}) \ \leftrightarrow \ \  \text{OH}^{*} \ + 1/2 \  \text{H}_{2}(\text{g}) .
\label{reaction1}
\end{align}
This reaction is of high interest to the electro-catalysis community for several reasons. First of all, the adsorption of OH molecules is believed to be responsible for the butterfly peak at $0.6$V$-0.85$V observed during cyclic voltammetry of water on Pt(111) \cite{Experimental_CV1, Experimental_CV2}. Secondly, as the ORR on Pt(111) takes place in the same potential range, it has been a subject of debate whether the co-adsorption of OH on Pt(111) has implications for the measured ORR activity \cite{Stamenkovic1, Koh1, Vukmirovic1, Wang1}. \newline
\indent In the context of modelling the reaction \eqref{reaction1}, the relevant physical relationship to determine is the coverage dependence of the OH adsorption energy. Early approaches such as those of Refs. \onlinecite{Rossmeisl, Rossmeisl1, Norskov, Tripkovic, Tripkovic1} determined this based on static, bilayer-like models of the interfacial hydroxyl-water layer, inspired by the structure that this has been hypothesized to adapt under ultra-high vacuum conditions (UHV) \cite{Ogasawara}. In particular, the conclusion from the study by Ref. \onlinecite{Rossmeisl1} was that the differential adsorption energy associated with reaction \eqref{reaction1} is essentially constant up to an OH coverage of $1/3$ML after which it exhibits a sharp jump, and they used this to explain the OH coverage profile measured in cyclic voltammetry experiments such as those of Refs. \onlinecite{Experimental_CV1, Experimental_CV2}. It has, however, been a subject of debate whether such UHV-inspired models provide a realistic description of reaction \eqref{reaction1} under the relevant operating conditions of electrochemical devices \cite{Gross, Gross1} and in a more recent study by Kristoffersen et al. \cite{Henrik} the adsorption energy associated with reaction \eqref{reaction1} was calculated based on room-temperature AIMD simulations with the liquid water layer explicitly included. 
The conclusion from this study was that the structure and energetics of the hydroxyls formed at the liquid water–Pt(111) interface were significantly different from the predictions of conventional bilayer-like models. It was, however, also noted that this conclusion was subject to statistical uncertainty due to the limited sampling capabilities of AIMD. Here we demonstrate how the machine learning framework of \cite{Mikkelsen2021} allows us to eliminate these sampling limitations and accurately determine the room temperature adsorption energy of \eqref{reaction1}. Due to its computational tractability, the machine learning framework furthermore allows us to investigate the structure and energetics of the adsorbed water-hydroxyl layers in large cell sizes, which would have been outside the scope of first-principles calculations. \newline
\indent The remainder of the paper is structured as follows. We first describe the computational details behind our methodology focusing on the construction of the NNP ensemble, the construction of the training database and the details of our MD simulations. We then present our results, where we first calculate the coverage dependent adsorption energy \eqref{reaction1} and compare to the AIMD study of Ref. \onlinecite{Henrik} as well as the low-temperature bilayer model. We then analyze the structure of adsorbed water-hydroxyl layers focusing on their hydrophobicity as well as the hydrogen bonding in these, and use the latter to explain the observed adsorption energetics.

\section{Methods}
\label{Methods}

\subsection{NNP ensemble}
\label{NNP}
Our NNP ensemble is a collection of NNPs obtained using the formalism proposed by Behler and Parinello \citep{Behler}. In this framework the potential energy surface (PES) of the system is written as a sum of atomic energies
\begin{align}
E = \sum_{i=1}^{N} E_{i}(\textbf{G}_{i}).
\label{E_NNP}
\end{align}
where $\textbf{G}_{i}$ denotes a vector of 2- and 3-body symmetry functions, which describe the local environment of each atom up to a cutoff radius, $R_{c}$ \citep{Behler2}.  
\newline 
Our motivation for using an ensemble of NNPs rather than a single one are the well known model- and data-based limitations of Behler-Parinello potentials \cite{Ceriotti, Mikkelsen2021} (see Section A for more details about this), and throughout the remainder of this paper we therefore use the spread of the ensemble predictions as a lower bound for the accuracy one can expect to achieve with our methodology. Concretely, the NNP ensemble consists of four NNPs all fitted using the RuNNer code \cite{Runner1}, where we employed the same set of symmetry functions as used in a previous NNP study of water on low-index Cu surfaces \cite{Natarajan}. The architecture of each NNP as well as the training and test errors are reported in Table \ref{table:NNP_table2}. The errors on the forces and energies are roughly a factor of 2 and 1.3 larger than the training and test errors reported in our previous study of the water/Pt(111) interface \cite{Mikkelsen2021}, though still comparable in magnitude to other training and test errors reported for similar studies using NNPs \cite{Natarajan}. The larger training and test errors can be understood as a result of a more long ranged character due to the presence of adsorbed hydroxyls. This is discussed in detail in Section A of the Supporting Information, where we have analyzed the data set and model limitations of our approach using learning curves as well as the locality test suggested by Deringer and Cs\'anyi \cite{Gabor}. The input and output files from the training of each NNP are publicly available and can be accessed via the DTU data repository \cite{OH_data}. 

\begin{table}[h!]
\centering
\begin{tabular}{ p{2.0cm}  p{4.0cm} p{4.0cm} p{4.0cm}  }
\toprule
& Architecture & $\text{E}_{\text{RMSE}} \ [\text{eV}/{\text{atom}}]$ & $\text{F}_{\text{RMSE}} \ [\text{eV}/{\text{Å}}]$\\
\midrule
NNP1 & 30-30 (s-s-l) & 0.0017(0.0016) & 0.084(0.084)\\
NNP2 & 30-25 (t-s-l) & 0.0016(0.0015) & 0.084(0.082)\\
NNP3 & 30-35 (t-s-l) & 0.0016(0.0015) & 0.084(0.088)\\
NNP4 & 30-40 (s-t-l) & 0.0016(0.0017) & 0.084(0.084)\\

\bottomrule
\end{tabular}
\caption{Summary of the architecture and training and test errors of the energies and forces for each NNP in our ensemble. For the architectures (first column), the number of nodes in each hidden layer is displayed as X-Y-Z etc. (i.e. 30-30 indicates two hidden layers with 30 nodes in each) and the activation functions used are indicated in parenthesis, where t, s and l are abbreviations for tanh, sigmoid and linear, respectively. For the training and test errors (second and third column) the first number is the training error and the second in parenthesis is the test error.}
\label{table:NNP_table2}
\end{table}

\subsection{Training database}
\label{Training_set}
Our NNP ensemble was trained on a dataset of 121\,377 structures consisting of a frozen $3 \times 4$ orthogonal Pt(111) slab with a water layer of $n_{\text{H}_{2}\text{O}} = 28, 27, .. , 20, 19, 18$ water molecules combined with $n_{\text{OH}} = 1, 2, ..., 8, 9, 10$ OH molecules, respectively.  All of these were obtained in an iterative manner starting from small data sets of $2000$ structures obtained from AIMD, which were then systematically expanded by performing MD simulations with preliminary NNP fits. This process was done to ensure that our training set densely covers the wide variety of structures encountered over long time scale MD simulations. The structures were set up using the Atomistic Simulation Environment (ASE) \cite{ASE}, and their energies and forces were determined using the PBE functional \citep{PBE} combined with the D3 van der Waals correction \citep{D3} as implemented in the Vienna ab initio Simulation Package (VASP) \cite{Vasp}. We employed a plane wave basis set with an energy cutoff of 350 eV and $2\times 2 \times 1$ k-points. The dataset is publicly available and can be accessed via the DTU data repository \cite{OH_data}.  \newline

\subsection{MD simulations}
\label{MD}
MD simulations with our NNPs were performed with the Lammps code \citep{lammps} using the interface provided by the n2p2 package \cite{n2p2, Singraber}. We performed constant temperature MD at 400K within the NVT ensemble using a Nosé-Hoover thermostat \citep{Nose-Hoover, Hoover1986} with a characteristic damping time of 50fs, and the Pt slab was kept frozen in all simulations. We have verified that the thermodynamic properties calculated below are consistent across different choices of thermostats with both weaker and stronger damping times. For a more thorough discussion and analysis of the effects of using different thermostats and coupling times for simulating liquid water on Pt(111) the reader is referred to the SI of \cite{Mikkelsen2021}. We use an elevated temperature for our simulations, as the PBE functional is known to overestimate the melting point of water \cite{PBE_meltingpoint}. Indeed, we found based on tests at different temperatures that a too low value in our MD simulations often led to periodic freezings of the water layer for several nanoseconds thereby hindering ergodic sampling - a problem which was further exacerbated by the presence of hydroxyls at the interface. We also note, that an elevated temperature has previously been employed in similar MD studies to mimic nuclear quantum effects on the oxygen distribution in water\cite{CalegariAndrade2020, Morrone2008}. Integration of the classical equations of motion were performed using Verlet integration \cite{Verlet} with a time step of 0.5fs, and we set hydrogen masses to $2$g/mol to allow for the use of a larger time step without affecting equilibrium statistical properties. For all simulation runs we also discarded the initial 100ps to ensure proper equilibration, and we employed an elastic wall in the upper part of the water layer to prevent the rare desorption of water molecules from the water layer onto the opposite side of our Pt slab.

\section{Results and discussion}

\label{Results}

\begin{figure*}[t!]
\includegraphics[width=\textwidth]{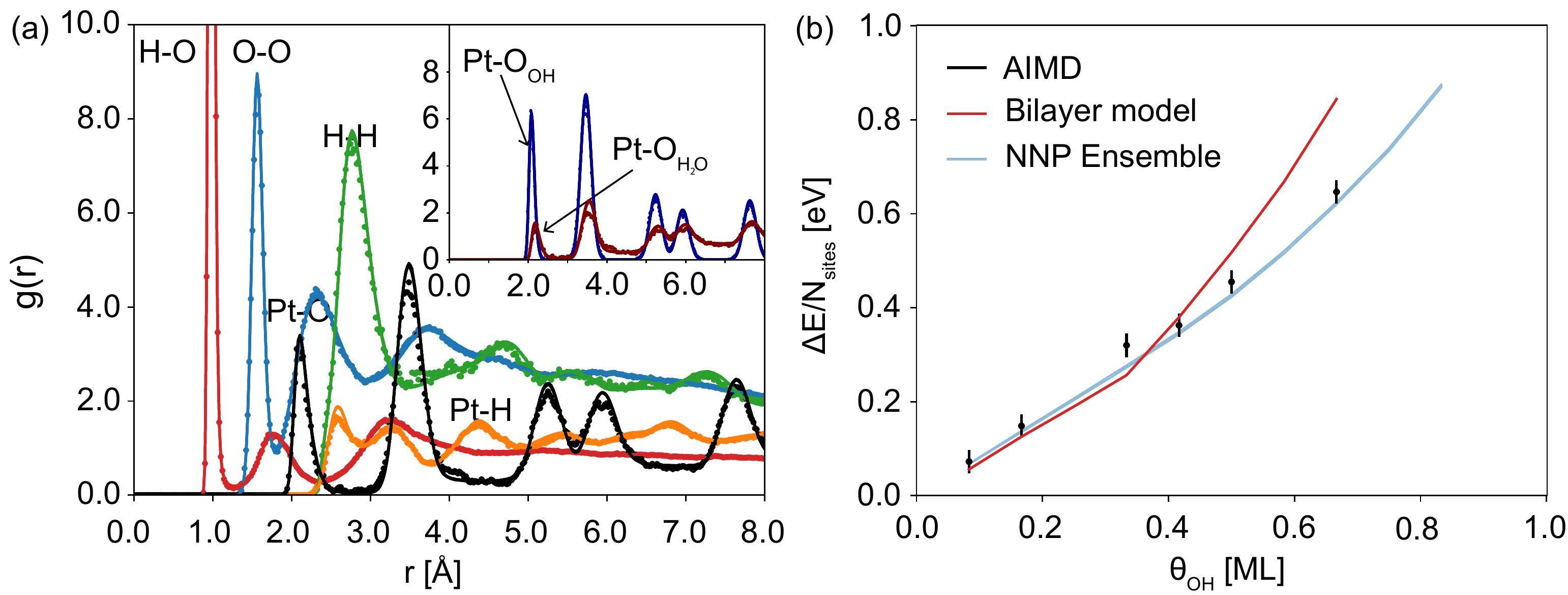}
\caption{(a) Radial distribution functions of selected atomic-type pairs as predicted by AIMD (points) and our NNP (lines). The width of the curves reflects the spread in the predictions of our NNP ensemble. Inset shows a zoom-in highlighting the Pt-O radial distribution function obtained using the top layer of the Pt(111) slab, which has been decomposed into partial RDFs associated with O atoms bound to OH (dark blue RDF) and water molecules (dark red RDF), respectively. (b) OH adsorption energies normalized by the number of Pt adsorption sites as predicted from AIMD (points with error bars reflecting statistical uncertainty) our NNP ensemble (blue line with thickness indicating the spread) and from the 0K bilayer model.}
\label{fig:RDFs_adsorption}
\end{figure*}

As a first assessment of the quality of our NNP ensemble we investigated its ability to describe the structure of the water-OH/Pt(111) interface. To do this, we conducted 10ns MD simulations with each NNP in our ensemble in a small unit cell with 12 Pt surface sites containing 4 adsorbed hydroxyls (1/3ML OH coverage) and compared selected radial distribution functions with those obtained from AIMD simulations in the same cell. As shown in Fig. \ref{fig:RDFs_adsorption} (a) there is good agreement between the predictions of our NNP ensemble (lines with thickness indicating ensemble spread) and AIMD (points). Furthermore, as shown in top-right inset, it is possible to define an adsorption layer of O atoms bound to the Pt surface similar to what was done in our previous work on the water/Pt(111) interface\cite{Mikkelsen2021}. However in this case, the surface-bound layer is seen to be composed of both hydroxyls and water molecules, as demonstrated by the decomposition of the Pt-O RDF into partials RDFs associated with hydroxyls (dark blue) and water molecules (dark red). The sharpness and position of the former reflects the well-known fact that OH molecules bind strongly to atop Pt sites in comparison with water molecules \cite{Michaelides_1, Karlberg2005}. 

\begin{table*}[ht]
\centering
\begin{tabular}{p{0.25\linewidth} p{0.05\linewidth} p{0.05\linewidth} p{0.05\linewidth} p{0.05\linewidth}p{0.05\linewidth} p{0.05\linewidth} p{0.05\linewidth} p{0.05\linewidth} p{0.05\linewidth} p{0.05\linewidth}}
\toprule
$\theta_{\text{OH}}$ [ML] & 0.083 & 0.167 & 0.250 & 0.333 & 0.417 & 0.500 & 0.583 & 0.667 & 0.750 & 0.833 \\
\midrule
$\Delta \text{E}_{\text{NNP1}}/n_{\text{OH}} \ $ [eV] & 0.82 & 0.83 & 0.83 & 0.83 & 0.84 & 0.86 & 0.89 & 0.94 & \text{N/A} & \text{N/A}   \\
$\Delta \text{E}_{\text{NNP2}}/n_{\text{OH}} \ $ [eV] & 0.81 & 0.82 & 0.82 & 0.82 & 0.83 & 0.85 & 0.88 & 0.93 & 0.98 & 1.05   \\
$\Delta \text{E}_{\text{NNP3}}/n_{\text{OH}} \ $ [eV] & 0.80 & 0.81 & 0.81 & 0.82 & 0.82 & 0.84 & 0.88 & 0.93 & 0.98 & 1.04   \\
$\Delta \text{E}_{\text{NNP4}}/n_{\text{OH}} \ $ [eV] & 0.81 & 0.82 & 0.83 & 0.83 & 0.83 & 0.85 & 0.89 & 0.94 & 0.99  & \text{N/A}    \\
$\Delta \text{E}_{\text{AIMD}}/n_{\text{OH}} \ $ [eV] & 0.87 & 0.89 & \text{N/A}  & 0.96 & 0.87 & 0.91 & \text{N/A} & 0.97 & \text{N/A} & \text{N/A}   \\
$\Delta \text{E}_{\text{Bilayer}}/n_{\text{OH}}$ [eV] & 0.67 & 0.75 & 0.76 & 0.77 & 0.91 & 1.04 & 1.15 & 1.27 & \text{N/A} & \text{N/A}   \\
\bottomrule
\end{tabular}
\caption{Normalized OH adsorption energies calculated from \eqref{adsorption_energy} with the bilayer model, AIMD as well as each NNP in our ensemble. In a few cases it was not possible to obtain the NNP-based adsorption energy at a particular coverage due to the underlying MD simulations being unstable.}
\label{table:adsorption_table}
\end{table*}
Having validated the quality of our NNP ensemble, we then focused on the coverage-dependent adsorption energy of reaction \eqref{reaction1}. To calculate this we employed the method used by Kristoffersen et al. \cite{Henrik}, who obtained this directly from the average total energies of their AIMD simulations using the following expression:
\begin{align}
\Delta E(n_{\text{OH}}) = \langle E_{(n_{\text{H}_{2}\text{O}} - n_{\text{OH}}) \cdot \text{H}_{2}\text{O} + n_{\text{OH}} \cdot \text{OH}}\rangle 
+ \frac{n_{\text{OH}}}{2} \cdot \langle E_{\text{H}_{2}} \rangle - \langle E_{n_{\text{H}_{2}\text{O}}  \cdot \text{H}_{2}\text{O}  }   \rangle 
\label{adsorption_energy}
\end{align}
here $\langle \cdot \cdot \rangle$ indicates a time-average over the different MD simulations, and $n_{\text{H}_{2}\text{O}}, n_{\text{OH}}$ denote the number of hydroxyls and water molecules present in the simulations, respectively. Finally, $E_{\text{H}_{2}}$ is the total energy of a gaseous H$_{2}$ molecule. \newline
The site-normalized OH adsorption energy profile calculated from \eqref{adsorption_energy} with our neural network potential ensemble is displayed in Fig. \ref{fig:RDFs_adsorption} (b) (opaque blue line), where we have included the predictions of the bilayer model\footnote{The calculations underlying these were performed in a 12 surface site unit cell employing the H-down bilayer structure suggested by Ogasawara et al. \cite{Ogasawara}. We have verified that the same conclusions are obtained for the convential bilayer structure suggested by Doering et al \cite{Doering}, where the hydrogen atoms point away from the slab.} (kinked red curve) and the AIMD calculated adsorption energies of Kristoffersen et al. \cite{Henrik} (black dots with error bars) for comparison. The NNP calculated adsorption energies were obtained from 10ns MD simulations conducted in $16.6$Å $\times \ 19.2$Å $\times \ 36$Å unit cells ($2$ $\times \ 2$ $\times \ 1$ replicas of the small unit cells used for Fig. \ref{fig:RDFs_adsorption} (a)) containing $n_{\text{OH}} = 0, 4, 8, 12, 16, 20, 24, 28, 32, 40$ OH molecules. Further details on the statistical convergence of these MD simulations are given in section B of the SI, where we show the use of nanosecond sampling times allows us to obtain internal energy averages, which agree within $\sim 0.025 \ \text{eV}$ for MD simulations started from different points in phase space. This is substantially better than what can be achieved with AIMD simulations such as those of Ref. \onlinecite{Henrik}, which, apart from being limited to small computational cells, are restricted to simulation times of $\sim 50$ps leading to sampling errors on the order of $\sim 0.2\text{eV}$ as reflected in the errorbars of Fig. \ref{fig:RDFs_adsorption} (b). \newline Evidently, our NNP-based results, which are within the statistical error \footnote{Note that deviations compared to Ref. \onlinecite{Henrik} might not only be due to sampling uncertainty, but also due the fact that we used a larger unit cell, that we kept the Pt slab frozen or that we used a larger simulations temperature.} of the AIMD results of Ref. \onlinecite{Henrik}, indicate a near-linear adsorption energy profile, where the cost of forming high coverage hydroxyl layers is significantly lowered compared to the bilayer model. In particular the kink at 1/3ML OH coverage, indicating a discontinous step in differential adsorption energy, is replaced by a weaker and more gradual increase in the slope of the adsorption energy profile. This conclusion may also be verified from Table \ref{table:adsorption_table}, which displays the adsorption energy per hydroxyl corresponding to the different curves in Fig. \ref{fig:RDFs_adsorption} (b). 

\begin{figure*}[t!]
\includegraphics[width=\textwidth]{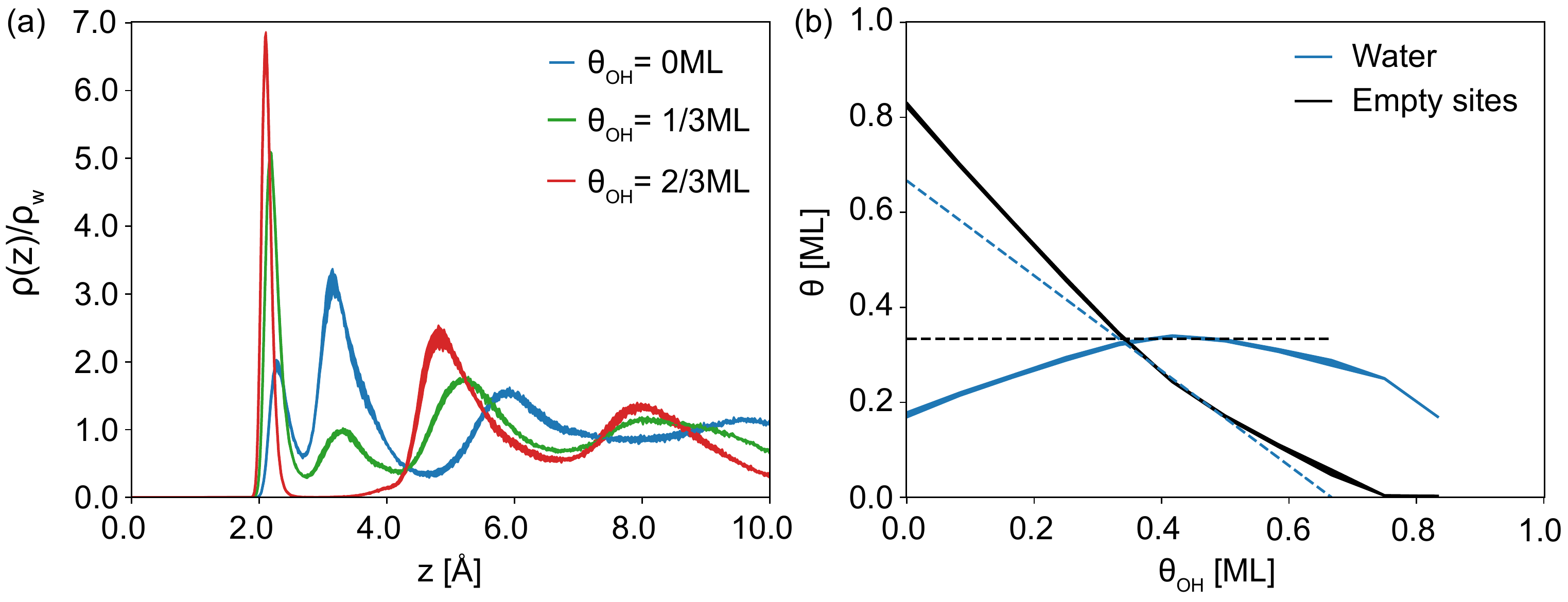}
\caption{(a) Planar averaged density away from the Pt surface for different OH coverages. (b) Coverage of water molecules and empty Pt surface sites as a function of OH coverage as predicted by the NNP ensemble (solid lines) and the bilayer model (dashed lines). The widths of the former indicate the spread of the ensemble predictions.}
\label{fig:Density_coverage}
\end{figure*}
What is the structure of adsorbed water-hydroxyl layers, and can the energetics observed above be understood from this? As a first step towards answering this we have analyzed the density profile at the interface as a function of hydroxyl coverage. The result of this is displayed in Fig. \ref{fig:Density_coverage} (a), where we plot the planar averaged density of water molecules as a function of distance from the Pt slab for OH coverages of $0\text{ML}, 1/3\text{ML}$ and $2/3\text{ML}$. In the case of no hydroxyl coverage, we obtain the characteristic double-peaked structure of the water/Pt(111) interface, which was analyzed in our previous work \cite{Mikkelsen2021}. In the presence of hydroxyls the same structure is observed, but it is clear that the presence of co-adsorbed hydroxyls pins the water molecules close to the Pt(111) surface leading to a depletion of the second density peak. This effect is further quantified in Fig. \ref{fig:Density_coverage} (b), which plots the coverage of water molecules as well as empty sites as a function of OH coverage (adsorbed water molecules are identified as those belonging to the first peak in the Pt-H$_2$O RDF of Fig. \ref{fig:RDFs_adsorption} (a)). Interestingly, the water coverage displays a maximum value around $\sim 0.4$ML, which reflects a balance between the strong Pt-OH attraction leading to a completely hydroxyl-covered Pt(111) surface at high coverage, and the water-OH attraction pinning the water molecules to the Pt(111) surface. 
\begin{figure*}[t!]
\centering
\includegraphics[width = \textwidth]{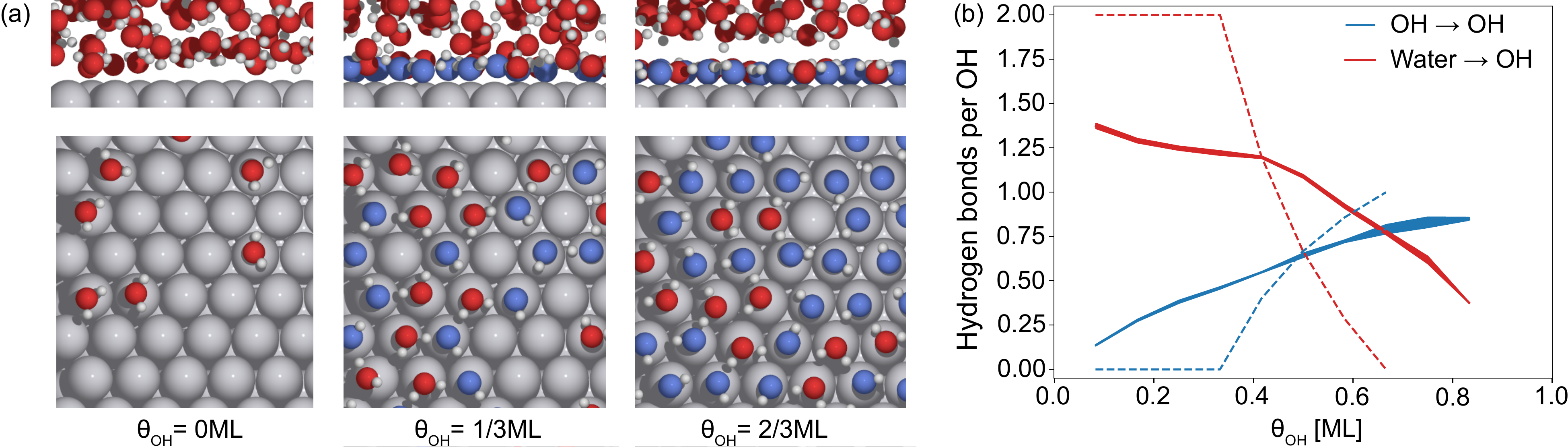}
\caption{(a) Snapshots of the interface at 0ML, 1/3ML and 2/3ML OH coverage. (b) Average number of hydrogen bonds donated from OH (blue) and water (red) to adjacent OH molecules as predicted by the NNP ensemble (solid lines) and the bilayer model (dashed lines). The widths of the former indicate the spread of the ensemble predictions.} 
\label{fig:H_bonds}
\end{figure*}
As shown in Fig. \ref{fig:H_bonds} (a), which shows typical MD snapshots for OH coverages of $0\text{ML}, 1/3\text{ML}$ and $2/3\text{ML}$, the findings above have interesting consequences for the structure of the interface. Evidently, the gradual removal of water molecules from the secondary peak in Fig. \ref{fig:Density_coverage} (a) leads to a $\sim$2Å depletion region between the water overlayer and adsorbed water-hydroxyl layer at high OH coverages, which suggests that the interface becomes hydrophobic with increasing hydroxyl coverage. This effect has also been observed experimentally in thermal desorption experiments \cite{Zimbitas2008}. To understand how all these observations can be used to explain the energetics of Fig. \ref{fig:RDFs_adsorption} (b), we have analyzed the hydrogen bonding in the adsorbed water hydroxyl layers using the geometric criterion employed by Natarajan et al \cite{Natarajan}. In particular we have focused on the number of hydrogen bonds donated by the adsorbed water molecules to the hydroxyls, as this has often been hypothesized to be the key stabilizing type of hydrogen bond in water hydroxyl layers on transition metal surfaces \cite{Schiros2007, Li2010, Forster2011}. As shown in Fig. \ref{fig:H_bonds} (b), the number of hydrogen bonds from water to OH (red) is rather flat up to $0.417$ML after which there is a rapid drop. This rapid drop is consistent with the increase in the normalized OH adsorption energies at high OH coverages observed in Table \ref{table:adsorption_table} and therefore offers a plausible explanation for the observed adsorption energetics.

\section{Conclusions}
We have investigated the room temperature structure and energetics of water-hydroxyl layers on Pt(111) by running nanosecond molecular dynamics simulations with an ensemble of neural network potentials. We first investigated the coverage dependent adsorption energy of OH molecules. A softly increasing OH adsorption energy profile was found, which is strikingly different from the conventional bilayer model but in good agreement with previous AIMD studies. We then analyzed the structure of the adsorbed water-hydroxyl layers, where we showed that the presence of co-adsorbed hydroxyls increases the hydrophobicity of interface resulting in the bulk overlayer being completely detached from the interface at high OH coverages. By analyzing the hydrogen bonding in the adsorbed water-hydroxyls we furthermore found that the number of hydrogen bonds donated from water to hydroxyl drops rapidly at large coverages, which we used to explain our calculated adsorption energies.

\begin{acknowledgement}

The authors acknowledge support from the Toyota Research Institute, the Villum Foundations through the research center V-Sustain (grant $\# 9455$) and the Department of Energy Conversion and Storage, Technical University of Denmark, through the Special Competence Initiative Autonomous Materials Discovery (AiMade).

\end{acknowledgement}

\begin{suppinfo}

\noindent \textbf{\large A. Model and data set limitations} \\ \\
Similar to the our previous study of the water/Pt(111) interface \cite{Mikkelsen2021}, we have analyzed the significance of long-range interactions and dataset limitations for our system using learning curves and the locality estimate suggested by Deringer and Cs\'anyi \cite{Gabor}. \newline 
The learning curve was determined by setting aside a fixed test set of 71\,377 structures chosen randomly from our database, and then incrementally expanding the training set using the remaining ones, while monitoring the error on the forces and energies. As shown in Fig. \ref{fig:Learning_Locality_WaterOH_Pt111} (a) the error in the forces and energies on the fixed test set drops rapidly as a function of training set size up to around only $\sim100$ structures after which the improvement is much more modest. This is similar to what was observed for the water/Pt(111) interface \cite{Mikkelsen2021}, though the decay is more gradual indicating that a larger number of chemically different structures are needed to accurately describe the energetics of the interface with hydroxyls present. The saturation values are also larger than those of reported in \cite{Mikkelsen2021}. Finally, we note that while the learning curve in Fig. \ref{fig:Learning_Locality_WaterOH_Pt111} (a) appears to suggest, that we could have done equally well with NNPs trained on much smaller training sets, this is in practice not true as the latter would give rise to instabilities during the NNP-based MD simulations. In fact, even with as large a training set as the one employed here, some of the trained NNPs still led to unstable MD simulations for certain OH coverages, where the atoms would drift into unreasonable parts of configuration space.  \newline
In the locality test proposed by Deringer and Cs\'anyi \cite{Gabor}, a central atom is chosen and every atom within a distance of $r_{\text{fix}}$ are then fixed while atoms outside are distorted randomly. By then monitoring the standard deviation of the force on the central atom, an estimate of the locality of the system is obtained. In our case we focused on the O, H and Pt atoms shown in Fig. \ref{fig:Learning_Locality_WaterOH_Pt111} (b) and performed calculations with a hydroxyl coverage of 1/3ML in a $2$ $\times \ 2$ $\times \ 1$ unit cell compared to the computational cell of the training structures ($1$ $\times \ 1$ $\times \ 1$ k-points were used for consistency) to be able to investigate values of $r_{\text{fix}}$ up to 8Å. Compared to the results of \cite{Mikkelsen2021} the long-range character is more pronounced with the presence of hydroxyls at the interface, as reflected in the slower decay of the force standard deviation and the fact that the curves saturate at values as large as $\sim 0.1\text{eV/Å}$. A plausible explanation for this more long-ranged behaviour could be that the interaction between co-adsorbed water and OH molecules is known to be of mainly electrostatic origin \cite{Karlberg2005}. We also note that the observation of a more long-ranged character, ties well with our previous comment of the higher training and test errors of our NNPs compared to those reported in our previous study\cite{Mikkelsen2021}.

\begin{figure*}[t!]
\includegraphics[width=\textwidth]{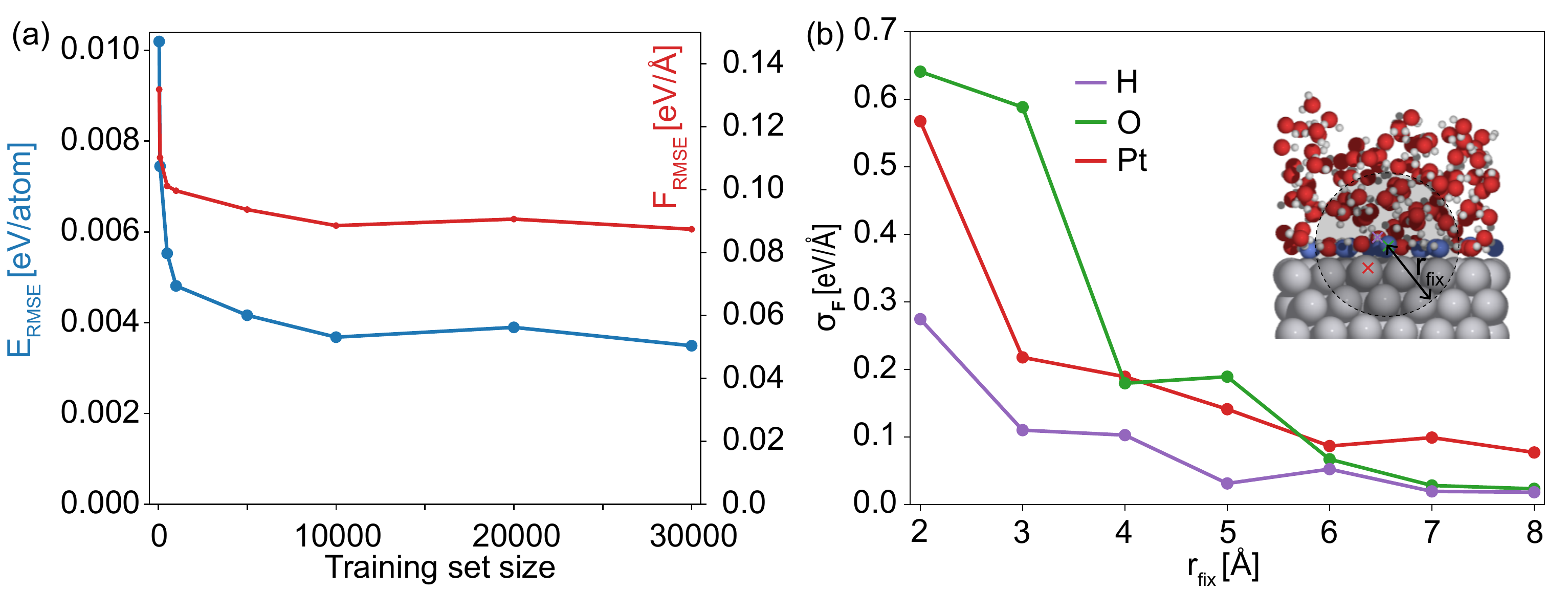}
\caption{(a) Learning curve displaying the error in the energies (blue line) and forces (red line) on a fixed test set as the training set is systematically expanded. (b) A measure of the locality of the water-OH/Pt(111) interface based on the procedure of Deringer and Cs\'anyi \cite{Gabor}. In the inset hydroxyls are colored blue, and the fixed O and H atom were chosen to belong to one of these.}
\label{fig:Learning_Locality_WaterOH_Pt111}
\end{figure*}

\newpage

\textbf{\large B. Obtaining accurately sampled internal energies} \\ \\ 
For every NNP in our ensemble the internal energy averages corresponding to the different OH coverages in \eqref{adsorption_energy} were determined from three 10ns MD simulations started from different structures. The average internal energy obtained from each of these runs are reported in Tables \ref{table:NNP1_averages}, \ref{table:NNP2_averages}, \ref{table:NNP3_averages}, \ref{table:NNP4_averages}, where it is clear that we obtain averages which agree within $\sim 0.025$eV for runs started from different points in phase space. To obtain the final values employed in \eqref{adsorption_energy} the averages corresponding to each coverage and NNP were averaged over the three independent runs.

\begin{table}[H]
\centering
\begin{tabular}{ p{2.0cm}  p{2.0cm} p{7.0cm} }
\toprule
$n_{\text{OH}}$ & $\theta_{\text{OH}}$ & $\langle E_{(n_{\text{H}_{2}\text{O}} - n_{\text{OH}}) \cdot \text{H}_{2}\text{O} + n_{\text{OH}} \cdot \text{OH}}\rangle $ [eV] \\
\midrule
0 & 0.0 & -2849.90, -2849.91, -2849.92 \\
4 & 0.083 & -2833.35, -2833.37, -2833.37 \\
8 & 0.167 & -2816.73, -2816.75, -2816.74 \\
12 & 0.250 & -2800.15, -2800.14, -2800.15 \\
16 & 0.333 & -2783.50, -2783.53, -2783.54 \\
20 & 0.417 & -2766.83, -2766.79, -2766.83 \\
24 & 0.500 & -2749.71, -2749.73, -2749.71 \\
28 & 0.583 & -2732.04, -2732.02, -2732.04 \\
32 & 0.667 & -2713.75, -2713.77, -2713.75 \\
36 & 0.750 & \text{N/A}  \\
40 & 0.833 & \text{N/A}  \\

\bottomrule
\end{tabular}
\caption{Average internal energies corresponding to different coverages and MD runs as predicted by NNP1. Values from the last two rows are absent as the NNP-based MD simulations were unstable for these two coverages.}
\label{table:NNP1_averages}
\end{table}

\begin{table}[H]
\centering
\begin{tabular}{ p{2.0cm}  p{2.0cm} p{7.0cm} }
\toprule
$n_{\text{OH}}$ & $\theta_{\text{OH}}$ & $\langle E_{(n_{\text{H}_{2}\text{O}} - n_{\text{OH}}) \cdot \text{H}_{2}\text{O} + n_{\text{OH}} \cdot \text{OH}}\rangle $ [eV] \\
\midrule
0 & 0.0 & -2849.69, -2849.69, -2849.69 \\
4 & 0.083 & -2833.15, -2833.17, -2833.17 \\
8 & 0.167 & -2816.60, -2816.62, -2816.58 \\
12 & 0.250 & -2800.05, -2800.02, -2800.07 \\
16 & 0.333 & -2783.47, -2783.46, -2783.48 \\
20 & 0.417 & -2766.82, -2766.81, -2766.85 \\
24 & 0.500 & -2749.76, -2749.80, -2749.78 \\
28 & 0.583 & -2732.09, -2732.04, -2732.14 \\
32 & 0.667 & -2713.75, -2713.71, -2713.81 \\
36 & 0.750 & -2694.95, -2694.92, -2694.96  \\
40 & 0.833 & -2674.97, -2674.96, -2674.96  \\

\bottomrule
\end{tabular}
\caption{Average internal energies corresponding to different coverages and MD runs as predicted by NNP2.}
\label{table:NNP2_averages}
\end{table}

\begin{table}[H]
\centering
\begin{tabular}{ p{2.0cm}  p{2.0cm} p{7.0cm} }
\toprule
$n_{\text{OH}}$ & $\theta_{\text{OH}}$ & $\langle E_{(n_{\text{H}_{2}\text{O}} - n_{\text{OH}}) \cdot \text{H}_{2}\text{O} + n_{\text{OH}} \cdot \text{OH}}\rangle $ [eV] \\
\midrule
0 & 0.0 & -2849.69, -2849.69, -2849.66 \\
4 & 0.083 & -2833.23, -2833.22, -2833.22 \\
8 & 0.167 & -2816.70, -2816.67, -2816.67 \\
12 & 0.250 & -2800.15, -2800.13, -2800.13 \\
16 & 0.333 & -2783.55, -2783.54, -2783.57 \\
20 & 0.417 & -2766.90, -2766.86, -2766.87 \\
24 & 0.500 & -2749.77, -2749.81, -2749.81 \\
28 & 0.583 & -2732.14, -2732.07, -2732.09 \\
32 & 0.667 & -2713.79, -2713.82, -2713.84 \\
36 & 0.750 & -2694.98, -2695.00, -2695.16  \\
40 & 0.833 & -2675.27, -2674.29, -2674.28  \\

\bottomrule
\end{tabular}
\caption{Average internal energies corresponding to different coverages and MD runs as predicted by NNP3.}
\label{table:NNP3_averages}
\end{table}

\begin{table}[H]
\centering
\begin{tabular}{ p{2.0cm}  p{2.0cm} p{7.0cm} }
\toprule
$n_{\text{OH}}$ & $\theta_{\text{OH}}$ & $\langle E_{(n_{\text{H}_{2}\text{O}} - n_{\text{OH}}) \cdot \text{H}_{2}\text{O} + n_{\text{OH}} \cdot \text{OH}}\rangle $ [eV] \\
\midrule
0 & 0.0 & -2849.88, -2849.87, -2849.87 \\
4 & 0.083 & -2833.35, -2833.37, -2833.38 \\
8 & 0.167 & -2816.76, -2816.74, -2816.78 \\
12 & 0.250 & -2800.16, -2800.12, -2800.13 \\
16 & 0.333 & -2783.53, -2783.48, -2783.51 \\
20 & 0.417 & -2766.88, -2766.83, -2766.89 \\
24 & 0.500 & -2749.84, -2749.81, -2749.83 \\
28 & 0.583 & -2732.11, -2732.14, -2732.10 \\
32 & 0.667 & -2713.80, -2713.74, -2713.75 \\
36 & 0.750 & -2694.98, -2694.91, -2694.90  \\
40 & 0.833 & \text{N/A}  \\

\bottomrule
\end{tabular}
\caption{Average internal energies corresponding to different coverages and MD runs as predicted by NNP4. Values from the last row are absent as the NNP-based MD simulations were unstable for this coverage.}
\label{table:NNP4_averages}
\end{table}
\end{suppinfo}

\bibliography{bibliography}

\end{document}